\documentclass{aa}
\usepackage{graphicx}
\usepackage{natbib}
\bibpunct{(}{)}{;}{a}{}{,} 
\begin{document}
\title{
\bf{The mineralogy, geometry and mass-loss history of \\ IRAS 16342-3814}
\thanks{Based on observations with ISO, an ESA project with instruments
        funded by ESA Member States (especially the PI countries: France,
        Germany, the Netherlands and the United Kingdom) with the
        participation of ISAS and NASA. The SWS is a joint project of
        SRON and MPE. Also based on observations obtained at the
	European Southern Observatory (ESO).}
	}

   \subtitle{}

   \author{C. Dijkstra
           \inst{1}
           \and
           L.B.F.M. Waters
	   \inst{1,2}
           \and
	   F. Kemper
	   \inst{1}
	   \and
	   M. Min
	   \inst{1}
	   \and
	   M. Matsuura
	   \inst{3}
	   \and
	   A. Zijlstra
	   \inst{3}
	   \and
	   A. de Koter
	   \inst{1}
	   \and
           C. Dominik
	   \inst{1}
          }

   \offprints{C. Dijkstra (dijkstra@astro.uva.nl)}

   \institute{Astronomical Institute, `Anton Pannekoek', University
              of Amsterdam, Kruislaan 403, NL-1098 SJ Amsterdam, The
              Netherlands
          \and
	      Instituut voor Sterrenkunde, Katholieke Universiteit Leuven,
              Celestijnenlaan 200B, B-3001 Heverlee, Belgium
	  \and 
	      Dept. of Physics, UMIST, Sackville Street, P.O. Box 88, 
	      Manchester M60 1QD, UK
             }

   \date{Received 7 October 2002 / Accepted 18 November 2002}

   \abstract{We present the $\rm{2-200{\ }\mu}$m {\em Infrared Space 
   Observatory} (ISO) spectrum and $\rm{3.8-20{\ }\mu}$m ISAAC and TIMMI2 
   images of the extreme OH/IR star IRAS 16342-3814. Amorphous silicate 
   absorption features are seen at $10$ and $\rm{20{\ }\mu}$m, together with 
   crystalline silicate absorption features up to almost $\rm{45{\ }\mu}$m. No
   other OH/IR star is known to have crystalline silicate features in 
   absorption up to these wavelengths. This suggests that IRAS 16342-3814 must
   have, or recently had, an extremely high mass-loss rate. Indeed, 
   preliminary radiative transfer calculations suggest that the mass-loss rate
   may be as large as $\rm{{10}^{-3}{\ }M_{\odot}{yr}^{-1}}$. The 
   $\rm{3.8{\ }\mu}$m ISAAC image shows a bipolar reflection nebula with a 
   dark equatorial waist or torus, similar to that seen in optical images 
   taken with the {\em Hubble Space Telescope} (HST). The position angle of 
   the nebula decreases significantly with increasing wavelength, suggesting 
   that the dominant source of emission changes from scattering to thermal 
   emission. Still, even up to $\rm{20{\ }\mu}$m the nebula is oriented 
   approximately along the major axis of the nebula seen in the HST and ISAAC 
   images, suggesting that the torus must be very cold, in agreement with the 
   very red ISO spectrum. The $\rm{20{\ }\mu}$m image shows a roughly 
   spherically symmetric extended halo, approximately $6''$ in diameter, which
   is probably due to a previous phase of mass-loss on the AGB, suggesting a 
   transition from a (more) spherically symmetric to a (more) axial symmetric 
   form of mass-loss at the end of the AGB. Using a simple model, we estimate 
   the maximum dust particle sizes in the torus and in the reflection nebula 
   to be $1.3$ and $\rm{0.09\mu}$m respectively. The size of the particles in 
   the torus is large compared to typical ISM values, but in agreement with 
   high mass-loss rate objects like AFGL 4106 and HD161796. We discuss the 
   possible reason for the difference in particle size between the torus and 
   the reflection nebula.
   }

   \maketitle
%

\section{Introduction}

\label{sect:introduction}

As a Main Sequence star with an initial mass between $1$ and about 
$\rm{8{\ }M_{\odot}}$ ages, it eventually evolves into an Asymptotic Giant 
Branch (AGB) star. An AGB star is a luminous cool giant star in which a 
dust driven wind expels matter from its envelope at a high rate
($10^{-7}$ to $\rm{10^{-4}{\ }M_{\odot}{yr}^{-1}}$). Mass-loss increases over 
time as the AGB star evolves. This eventually ends in an episode of extremely 
high mass-loss, the so-called superwind (SW) phase, in which the mass-loss 
rate, $\dot{M}$, exceeds $\rm{10^{-5}{\ }M_{\odot}{yr}^{-1}}$. The SW rapidly 
strips the star of virtually its entire envelope. The high mass-loss during 
the AGB phase (the total AGB phase lasts some $\rm{10^{6}{\ }yr}$) creates a 
circumstellar envelope which may completely obscure the star. When the SW 
phase stops, the central star starts to shrink in radius at constant 
luminosity, getting hotter over time. The star is now said to be in the 
post-AGB phase. The old circumstellar envelope, which initially obscures the 
central star, becomes optically thin at optical wavelengths after 
$\rm{\sim{100-1000{\ }yr}}$ because it is still expanding. Depending on the 
rate at which the central star evolves, the star may start to photoionize the 
material that was ejected during the AGB phase, creating a planetary nebula 
(PN). In this case the phase between the end of the AGB and the birth of the 
PN is called the proto-planetary nebula (proto-PN) phase.

In this paper we study the proto-PN IRAS 16342-3814 (hereafter: IRAS\,16342; 
also OH 344.1+5.8 or "the water-fountain nebula"). The evolutionary 
status of IRAS\,16342 as a proto-PN is examined in detail by 
\citet[][ hereafter: SAH99]{1999ApJ...514L.115S}. They report the
presence of an asymmetrical bipolar reflection nebula in 
{\em Hubble Space Telescope} (HST) images at $0.55$ and $\rm{0.80{\ }\mu}$m. 
The two lobes of the nebula, where the eastern lobe is significantly redder 
than the western lobe, are separated by a dark equatorial waist. The nebula is 
illuminated by starlight escaping through the polar holes in this `optically 
thick, dense, flattened cocoon of dust'. The central star is completely 
obscured by the dust cocoon (hereafter called `the torus'). The individual 
symmetry axes of the lobes do not coincide, although their orientation and 
location with respect to the dark waist are centrally point symmetric. OH 
maser line data at 1612, 1665 and 1667 MHz overlaid on the $\rm{0.80{\ }\mu}$m 
image suggest that the nebular axis is tilted such that the western lobe is 
nearest to us. Faint halos are seen around the lobes. The halo regions are 
illuminated by scattered light from the lobes, and possibly by heavily 
reddened, direct starlight (SAH99). 

IRAS\,16342 has been observed as part of a mid-infrared 
\citep{1999ApJS..122..221M} and optical imaging survey 
\citep{2000ApJ...528..861U}. In the mid-infrared survey IRAS\,16342 was 
observed at $9.8$, $12.5$, $17.8$ and $\rm{20.6{\ }\mu}$m.
At these wavelengths the star was classified as a core/elliptical 
source, i.e. a source with an unresolved, very bright core that is surrounded 
by a lower surface brightness elliptical nebula. At all mid-infrared 
wavelengths the source was found to extend northeast-southeast with the same 
position angle for all wavelengths. 

Early studies on IRAS\,16342 revealed the presence of extremely high velocity 
outflows in both $\rm{H_{2}O}$ ($\rm{{\sim}130{\ }km{\ }s^{-1}}$) and OH 
($\rm{{\sim}70{\ }km{\ }s^{-1}}$) maser line emission 
\citep{1987A&A...173..263Z,1988ApJ...329..914L}. 
Likkel \& Morris interpreted the $\rm{H_{2}O}$ masers as streams or
clumps of molecular gas forced out at the polar axes, and believed the OH 
masers to originate from lower velocity material at intermediate latitudes. 
\citet{1992A&A...256..581L} suggested a binary star model for IRAS\,16342, 
as this could account for both the bipolarity and the high outflow velocities
of the masers. 

Radiative transfer calculations on IRAS\,16342 have been performed by
\citet{1996A&A...305..878G} and \citet{2001MNRAS.322..280Z}. 
The latter authors found an inner radius of $1''$ and an outer radius
of $1.5''$ for the torus, in agreement with the HST observations. The mass of 
the torus was found to be $\rm{0.1{\ }M_{\odot}}$. The radiative transfer 
calculations by both groups suggest the presence of a (very cool) outer shell.

In this paper we present the $\rm{2-200{\ }\mu}$m {\em Infrared Space 
Observatory} (ISO) spectrum and infrared ISAAC and TIMMI2 images ($3.8$ to 
$\rm{20{\ }\mu}$m) of IRAS\,16342. The aim of this paper is to study the 
mineralogy, morphology and mass-loss history of IRAS\,16342. The paper should 
also add to our understanding of proto-PNe in general and to the injection of 
matter into the ISM by AGB stars. The latter is important since AGB stars are 
among the main `factories' of interstellar dust in galaxies. In Sect. 2 the 
data and data reduction procedures are described. In Sect. 3 we present and 
discuss the ISO spectrum. Based on the spectrum, we study the mineralogy and 
mass-loss history of IRAS\,16342. The ISAAC and TIMMI2 images of IRAS16342 are 
presented in Sect. 4 and will be used to study the morphology of IRAS\,16342
and set limits on the grain sizes in the torus and the lobes. The conclusions 
are listed in Sect. 5.


\section{Data and data reduction}

\subsection{ISO data}

\label{isodatareduction}

   \begin{table}[t]
      \caption[]{The IRAS fluxes for IRAS\,16342 compared to those derived from
 		the ISO data. From left to right we list the wavelength; the 
		Point Source Catalog (PSC) IRAS flux; the IRAS flux derived 
		from the ISO spectrum. For more details see Sect. 
		\ref{isodatareduction}.}
         \label{tab:fluxes}
     $$ 
         \begin{array}{ccc}
            \hline
            \noalign{\smallskip}
	\lambda &\rm{PSC{\ }IRAS}&\rm{ISO{\ }flux}\\
	(\rm{\mu}m) &\rm{flux{\ }(Jy)}&\rm{(Jy)}\\
            \noalign{\smallskip}
            \hline
            \noalign{\smallskip}
	    12	& 16.2  &12.0\\ 
	    25	& 199.8 &170.3\\	
	    60	& 290.2 &292.5\\
	    100	& 139.4 &165.3\\
            \noalign{\smallskip}
            \hline
         \end{array}
     $$ 
   \end{table}

The $\rm{2-200{\ }\mu}$m spectrum was obtained with ISO (Kessler et al. 1996), 
using the {\em Short Wavelength Spectrometer} 
\citep[SWS;][]{1996A&A...315L..49D} for the $\rm{2-45{\ }\mu}$m region and the 
{\em Long Wavelength Spectrometer} \citep[LWS;][]{1996A&A...315L..38C} for 
the $\rm{45-200{\ }\mu}$m region. The two spectra were reduced independently. 
For {\bf both} the SWS and {\bf LWS} reduction, we used pipeline version 
OLP 10 as the starting point. In both cases, individual subbands 
were cleaned from glitches, flatfielded (to the mean of all detectors), 
sigmaclipped (using the default values $\sigma=3$ for SWS and $\sigma=2.5$ 
for LWS) and rebinned (${\lambda}/{\Delta}{\lambda}=500$ for SWS and 
${\Delta}{\lambda}=\rm{0.068{\ }\mu}$m for LWS). The subbands were scaled (to 
band 1 and/or 2 for SWS and detector 5 for LWS) to obtain a continuous 
spectrum. Finally, the LWS spectrum was scaled to the SWS spectrum to get 
one continuous spectrum. Without scaling, the flux levels of the two 
spectrographs matched at $\rm{45{\ }\mu}$m to within $12{\%}$. For LWS there 
was an off source spectrum available to correct for interstellar flux 
contributions. The flux levels in this spectrum were low 
(${\leq}10{\ }\rm{Jy}$), showing that the on-source spectrum has no 
significant background contribution. We have compared our reduction with the 
available IRAS data. Between $14$ and $\rm{25{\ }\mu}$m, the ISO spectrum 
overlaps the IRAS Low Resolution Spectrum (LRS) to within $5{\%}$, while 
between $12$ and $\rm{14{\ }\mu}$m the IRAS LRS flux is about $25{\%}$ higher 
than the ISO flux. The IRAS Point Source Catalog (PSC) fluxes match those 
derived from the ISO data to within $26{\%}$ (see Table \ref{tab:fluxes}).

\subsection{ISAAC data}

\label{sect:isaacdata}

The IRAS\,16342 $\rm{3.8{\ }\mu}$m (L-band) image was obtained with the 
{\em Infrared Spectrometer And Array Camera} (ISAAC)
\footnote{http://www.eso.org/instruments/isaac/index.html} at the {\em Very 
Large Telescope} (VLT) at Paranal, Chile, on August 6, 2000, using the Santa 
Barbara Research Center Aladdin 1024$\times$1024 pixel InSb array. The pixel 
scale of this detector is 0.071 $''$ per pixel. The total field of view is 
72$''$ $\times$ 72$''$. The L-band filter has a central wavelength of 
3.78\,$\rm{\mu}$m and a width of 0.58\,$\rm{\mu}$m. The optical seeing during 
the observations varied between 1.2$''$ and 2.0$''$.

The image was obtained using a 10$''$ chop throw (perpendicular to the 
long axis of the source), while nodding was performed. In addition to the 
chopping, the source position was also jittered. The total integration 
time was 5.0 minutes, not considering the negative image in the chopped image.
Only the positive image was used, since the PSF of the negative image was 
systematically wider. We used two photometric calibration stars: Y3501 and 
HR6736. The L-band magnitudes of these stars were measured by MSSSO photometry
\citep{1994PASP..106..508M} and \citet{1996A&AS..119..547V}, respectively. As 
a reference for the point spread function (PSF), a nearby star, 
USNO 450$-$24728679 ($m_{\rm{L}}=7.0{\ }\rm{magn}$) was used. The obtained PSF
was used to deconvolve the ISAAC image of IRAS\,16342. The FWHM of the PSF was
calculated to be 0.35$''$. The reduction was done using eclipse package 
version 4.1.2 \citep{1997TheMessenger...87S}, and DAOPHOTO under IDL. 

\subsection{TIMMI2 data}

\label{sect:timmi2data}

IRAS\,16342 images at $7.9$ (narrow band), $9.8$ (narrow band), $10.6$ (broad 
band) and $\rm{20.0{\ }\mu}$m (broad band) were observed with the {\em Thermal 
Infrared Multi Mode Instrument} (TIMMI2) at the ESO 3.60m telescope at La 
Silla, Chile, on August 11, 2001, using a Raytheon 240$\times$320 pixel AsSi 
array. The pixel scale of this detector is 0.202$''$ per pixel. The recorded 
optical seeing during the observations varied between 0.92$''$ and 2.0$''$.

The images were obtained using a 15$''$ chop throw in the north-south 
direction while nodding was performed. As a reference for the PSF, 
HD 124897 ($={\alpha}$ Boo) and HD 178345 were used. We combined individual
chopped images into one image to improve the signal-to-noise. This was
done by shifting the positive and negative images in each chopped
frame to a prescribed position and then adding or subtracting
them. Frames of poor quality were omitted. Finally, we deconvolved the 
resulting image, using either HD\,124897 or HD\,178345 as PSF. Due to 
non-photometric weather during the observations, the absolute flux calibration
of the TIMMI2 images proved difficult and the results uncertain. Therefore, 
further attempts to do an absolute flux calibration on the images were not 
made. Moreover, we are mainly interested in the morphology of IRAS\,16342 at 
different wavelengths, which does not require an accurate (absolute) 
calibration of the flux. We will use relative fluxes in the individual images 
only.

\section{The ISO spectrum}

\subsection{Description of the spectrum}

\label{sect:spectrum}

\begin{figure*}
  \caption{The $\rm{2-200{\ }\mu}$m ISO spectra of IRAS\,16342-3814 (top 
	   panel), the OH/IR star AFGL 5379 (second panel) and the post-AGB 
	   star HD 161796 (third panel). Adopted continua for the spectra are 
	   given (smooth solid lines). The lower panel shows the continuum 
	   subtracted spectra of IRAS\,16342 (solid line) and HD 161796 
	   (dashed line). For a discussion on see Sect. \ref{sect:spectrum}.}
  \hspace{-0.55cm}\rotatebox{90}{\resizebox{14cm}{!}{\includegraphics{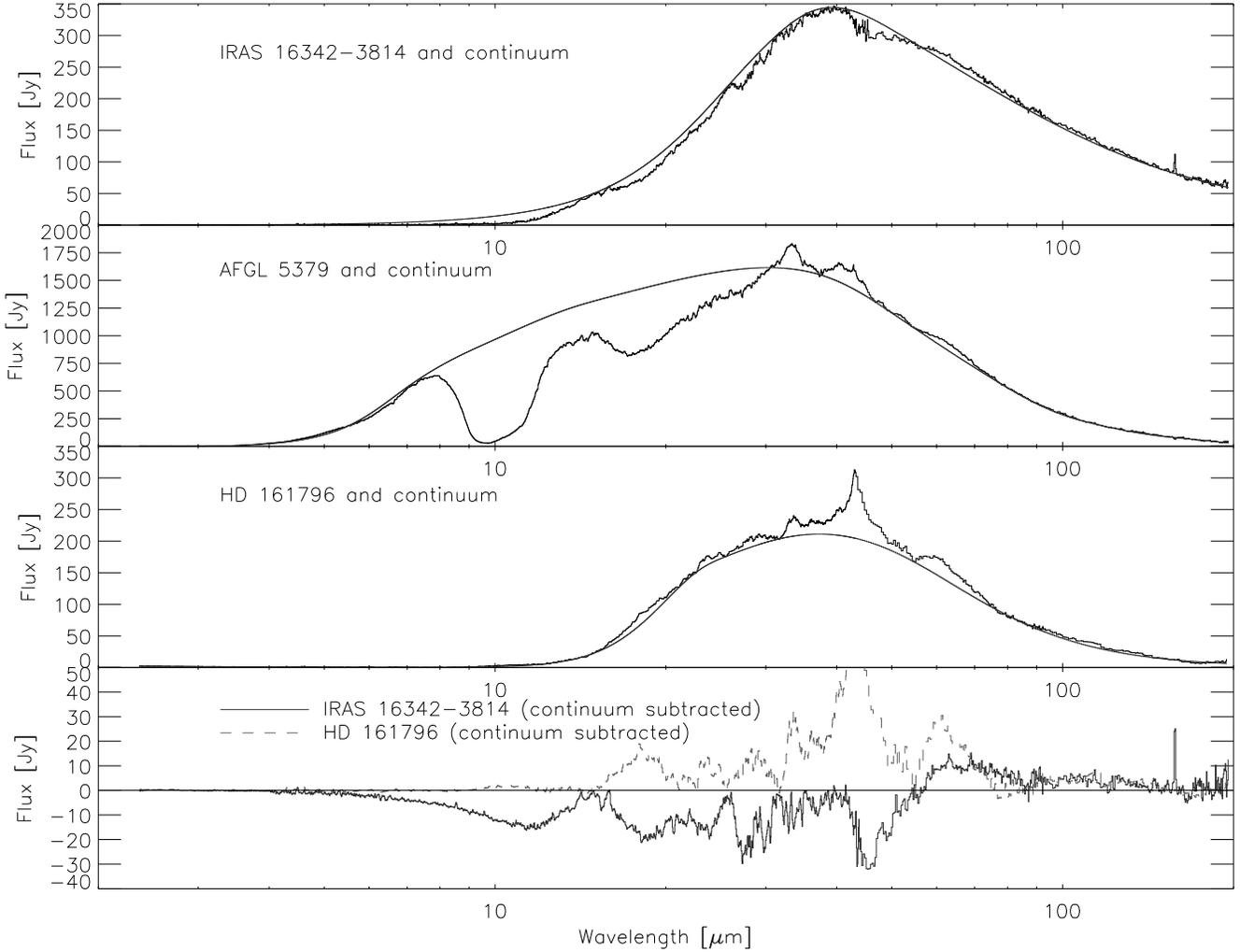}}}
  \label{fig:ISOspectrum}
\end{figure*}

In the top panel of Fig. \ref{fig:ISOspectrum} we present the 
$\rm{2-200{\ }\mu}$m ISO spectrum of IRAS\,16342 (solid line). A possible 
continuum for the spectrum, which we will discuss below, is also given (smooth
solid line). The second panel shows the spectrum of the OH/IR star AFGL 5379, 
studied in detail by \citet{1999A&A...352..587S}. The continuum for AFGL 5379 
is taken from that paper. The $10$ and $\rm{20{\ }\mu}$m amorphous 
silicate features are clearly visible. These absorption features are observed 
in many AGB stars, their strength depending on the optical depth of the dust, 
which is directly related to the mass-loss rate. The absorption features 
become stronger, and the spectrum redder, as $\dot{M}$ increases (e.g. 
Bedijn, 1987). IRAS\,16342 also shows the amorphous silicate features. The one
at $\rm{20{\ }\mu}$m was already detected in the IRAS LRS spectrum and fitted 
by \citet{2001MNRAS.322..280Z}. Compared to AFGL 5379, the amorphous silicate 
features in the spectrum of IRAS\,16342 might seem to be absent at first 
glance! The spectrum of IRAS\,16342-3814 is that of an extremely reddened 
OH/IR star. The third panel in Fig. \ref{fig:ISOspectrum} shows the spectrum 
of HD 161796. HD 161796 is another post-AGB object which has been studied in 
detail by \citet{2002A&A...389..547H} and \citet{2002ApJ...571..936M}. The 
continuum for HD 161796 is taken from \citet{2002A&A...389..547H}. 

A first inspection of the spectrum of IRAS\,16342 suggests the presence of the 
characteristic features of crystalline silicates and crystalline water ice, 
though up to $\rm{45{\ }\mu}$m they appear in absorption instead of in 
emission. To illustrate this, the continuum shown in the top panel of Fig. 
\ref{fig:ISOspectrum} has been drawn accordingly (smooth solid line). 
Following \citet{2002A&A...382..184M}, we represented the continuum by a 
spline-fit and required it to be smooth and to maximize the
continuum flux in both $F_{\nu}$ and $F_{\lambda}$. We emphasize that this 
continuum is used only to enhance the visibility of sharp features in the 
spectrum and has no physical meaning. In the lower panel of Fig. 
\ref{fig:ISOspectrum} we plot the continuum subtracted spectra of IRAS\,16342 
(solid line) and HD 161796 (dashed line). The continuum subtracted spectrum of
HD 161796 is dominated by crystalline silicate and crystalline water ice 
emission features \citep{2002A&A...389..547H}. A comparison of the suggested 
absorption features in IRAS\,16342 with the emission features in HD 161796 
shows an impressive agreement, indeed identifying the IRAS\,16342 features as 
due to crystalline silicates and water ice, and justifying our original choice
of the continuum. At wavelengths longwards of approximately $\rm{45{\ }\mu}$m 
the features in the spectrum of IRAS\,16342 are again in emission. Close to 
the peak of the SED, where the transition between absorption and emission 
occurs, the shape of the spectral features is less certain.

\subsection{The mineralogy of IRAS\,16342}

\begin{figure}
  \caption{The $\rm{15-55{\ }\mu}$m and $\rm{40-100{\ }\mu}$m continuum 
           subtracted spectrum of IRAS\,16342 together with the materials 
	   identified in it. The mass absorption coefficients for forsterite 
	   (F), clino-enstatite (E) and diopside (D), together with the 
	   absorbance spectrum of crystalline water ice (H2O) are also shown. 
	   The continuum in the mass absorption coefficient and absorbance 
	   curves has been subtracted in order to enhance the spectral 
	   features of the materials. Identifications of spectral features are
	   given where possible. Note that arbitrary scaling and offset values
	   have been applied to the laboratory data in order to reveal the 
	   spectral structures in it as good as possible. For a given 
	   material, the values are not necessarily the same in both panels.}
  \rotatebox{270}{\resizebox{13cm}{!}{\includegraphics{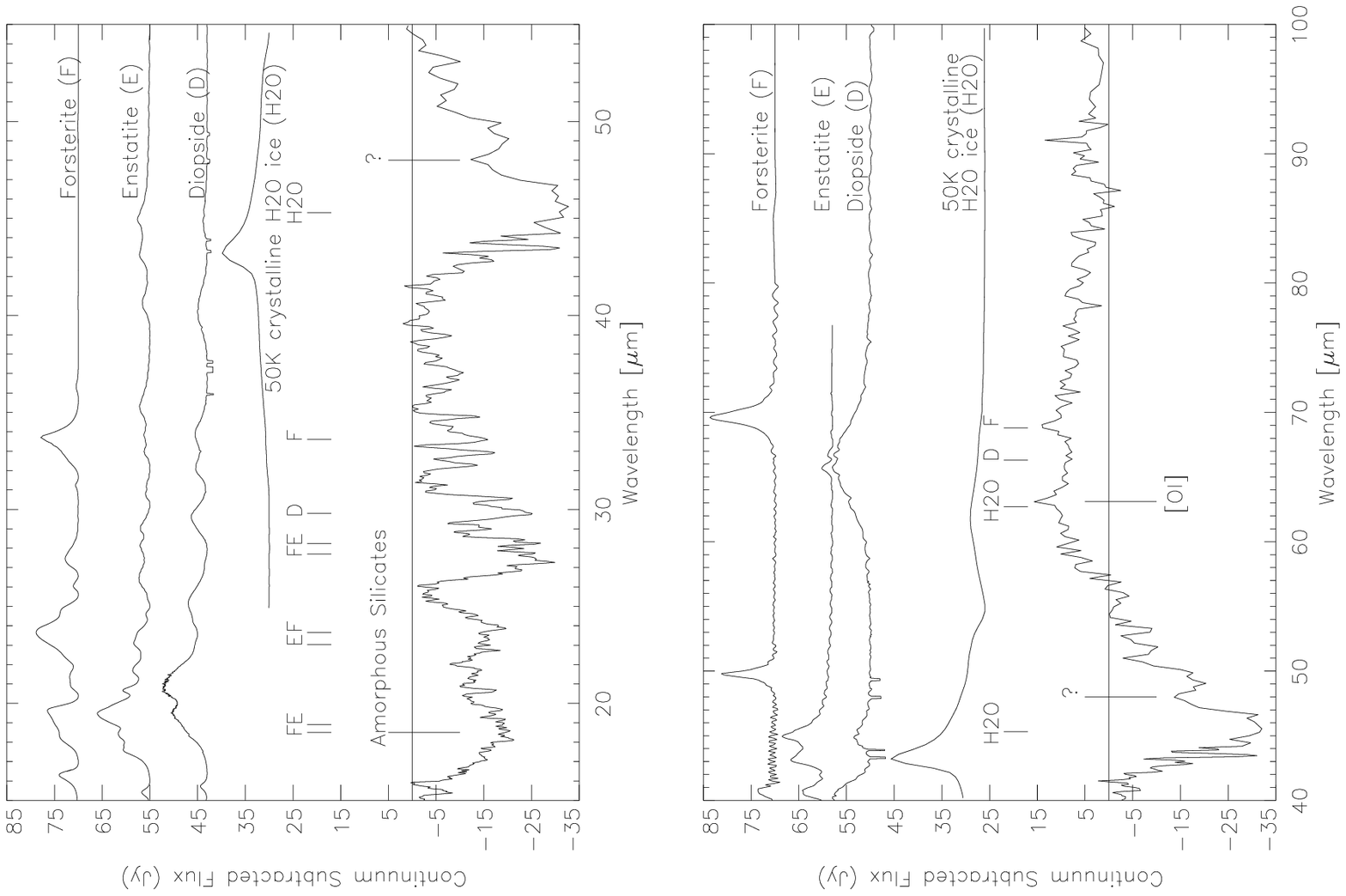}}}
  \label{fig:continuumsubtractedISOspectrum}
\end{figure}

   \begin{table}
      \caption[]{A list of the wavelength positions of the spectral
		features found in the $\rm{2-200{\ }\mu}$m spectrum of IRAS 
		16342-3814. Where available an identification of a feature is 
		given. Note that the amorphous silicates are positioned at 
		approximately 10.9 $\rm{\mu}$m, which is at a longer 
		wavelength than the normal 9.7 $\rm{\mu}$m. This difference is
		probably due to saturation of the spectrum in the 
		10 $\rm{\mu}$m region, making it difficult to determine a 
		suitable continuum in this region and affecting the shape and 
		position of the feature in the continuum subtracted spectrum.
		}
         \label{tab:mineralogy}
     $$ 
         \begin{array}{lp{0.5\linewidth}p{0.2\linewidth}}
            \hline
            \noalign{\smallskip}
            \lambda ({\mu}m)&Identification&Remarks  \\
            \noalign{\smallskip}
            \hline
            \noalign{\smallskip}
	    10.9\pm0.4    &Amorphous Silicates       &absorption \\
	    18.5\pm0.1    &Amorphous Silicates+      &absorption \\
	      		  &Forsterite+Clino-Enstatite$^{\mathrm{\dag}}$(?)&absorption\\
	    23.5\pm0.1    &Forsterite+Clino-Enstatite$^{\mathrm{\dag}}$(?)&absorption\\
	    27.6\pm0.1    &Forsterite+Clino-Enstatite$^{\mathrm{\dag}}$(?)&absorption\\
	    29.8\pm0.1    &Diopside(?)               &absorption \\
	    30.6\pm0.1    &? (Unidentified)          &absorption \\
	    33.6\pm0.1    &Forsterite                &absorption \\
	    36.1\pm0.1    &? (Unidentified)          &absorption \\
	    36.9\pm0.1    &? (Unidentified)          &absorption \\
	    45.3\pm0.5    &Crystalline Water Ice+    &absorption \\
	                  &Crystalline Silicates(?)  &           \\
	    48.0\pm0.1    &? (Unidentified/Artifact?)&emission   \\
	    62.7\pm0.3    &Crystalline Water Ice+    &emission   \\
                	  &Diopside                  &broad      \\
	    63.1\pm0.1    &[OI]                      &emission   \\
	    68.8\pm0.1    &Forsterite                &emission   \\
	    158.0\pm0.1   &[CII]                     &emission   \\
            \noalign{\smallskip}
            \hline
         \end{array}
     $$ 
	\begin{list}{}{}
	\item[$^{\mathrm{\dag}}$] By analogy with HD 161796 we tentatively
	identify clino-enstatite in the spectrum of IRAS\,16342. 
	\end{list}
   \end{table}

Table \ref{tab:mineralogy} lists the wavelength positions and identifications 
of the features found in the spectrum of IRAS\,16342. We identify 
amorphous silicates, forsterite ($\rm{Mg_{2}SiO_{4}}$), 
diopside ($\rm{MgCaSi_{2}O_{6}}$), crystalline water ice, and possibly 
clino-enstatite ($\rm{MgSiO_{3}}$) as important solid state components of the 
circumstellar matter around this object. We also find [OI] and [CII] lines, 
indicating low density gas. The off source spectrum suggests that the [CII] 
line is largely interstellar, although a circumstellar contribution (up to 
$25{\%}$ in peak strength) may still be present. The [OI] line seems to be 
interstellar. After subtraction of the off source spectrum, in which the [OI] 
line is clearly seen, there is no clear evidence for a residual circumstellar 
component. 

Fig. \ref{fig:continuumsubtractedISOspectrum} shows part of the 
continuum subtracted spectrum of IRAS\,16342 in more detail, together with the 
mass absorption coefficients for forsterite, clino-enstatite and diopside 
\citep{1999SpectraKoike}. Also shown is the absorbance of crystalline water 
ice measured at 50 K \citep{1994MNRAS.271..481S}. The wavelength positions of 
the identified forsterite, diopside and crystalline water ice
features are indicated with short vertical lines. We also indicate the 
positions of clino-enstatite bands. Due to blending with forsterite, and given
that the bands are fairly weak, we cannot be certain that clino-enstatite is 
present. However, in most evolved stars both forsterite and enstatite are 
found together \citep{2002A&A...382..222M}, and we suggest this is also the 
case in IRAS 16342. The $\rm{18.5{\ }\mu}$m amorphous silicate feature, the 
[OI] line and the unidentified $\rm{48{\ }\mu}$m feature 
\citep[e.g.][]{2002A&A...382..222M} are also indicated.

The solid state materials found in the circumstellar environment of IRAS\,16342
are also found around other (high mass-loss), oxygen rich, evolved stars 
\citep[e.g.][]{2000A&A...363.1115K,2002A&A...382..222M}. The very red ISO    
spectrum suggests that the bulk of the material around IRAS\,16342 must be     
cold. A low temperature may also be implied by the $\rm{69{\ }\mu}$m feature 
of forsterite. This intrinsically weak feature is detected either when
the forsterite is cold or very abundant. The feature at $\rm{48{\ }\mu}$m has 
been found in the spectra of many sources \citep[e.g.][]{2002A&A...382..222M}. 
Its nature is unclear at this time. FeSi has been proposed as a possible      
carrier \citep{2000A&A...357L..13F}, but this identification remains debatable
\citep{2002A&A...382..222M}. The $\rm{48{\ }\mu}$m feature is seen in 
emission, superposed on the $\rm{45{\ }\mu}$m feature of crystalline water 
ice, which is in absorption. If real, this suggests that the carrier of the 
feature has a different spatial distribution than the crystalline water ice, 
otherwise both materials would be seen in absorption. An alternative 
explanation is that the feature is an artifact in the responsivity function of
the ISO LWS detector. More study will be needed to explore this issue.
      
\subsection{The Mass-loss history of IRAS\,16342}

\label{sect:masslosshistory}

No other OH/IR star known to us displays a spectrum reddened to the extent as 
is seen in IRAS\,16342. Also, no other OH/IR star is known to have crystalline
silicate features in absorption at wavelengths up to $\rm{45{\ }\mu}$m. Though 
\citet{1999A&A...352..587S} showed that crystalline silicate absorption 
features are present in OH/IR stars, they did not detect them longwards of 
$\rm{25{\ }\mu}$m. The behaviour of the spectrum and the crystalline features 
in it is the result of a very high optical depth in the dust shell, which 
suggests that the star must have an extremely large mass-loss rate.

We modeled the spectrum of IRAS\,16342 using the radiative transfer programme 
{\sc modust} \citep{2001PhDT..........B}. We did not try to make a detailed 
fit of the spectrum, but rather wanted to get an order of magnitude estimate 
on the mass-loss rate. We assumed an M9III AGB central star 
($T_{\rm{eff}}{\sim}\,2670{\ }\rm{K}$ and $R_{\star}=372{\ }\rm{R_{\odot}}$)
surrounded by a spherical circumstellar envelope (composed of olivine and iron
particles, following \citet{2002A&A...384..585K}) created by mass-loss from 
the central star. We derived 
$\rm{\dot{M}\approx{10}^{-3}{\ }M_{\odot}{yr}^{-1}}$. In this case the best 
match between the peak position of the synthetic spectrum and that of the data
was obtained. Also, the lack of near infrared ($\rm{2-10{\ }\mu}$m) flux was 
best modeled in this case. 

The crystalline silicate absorption features in the ISO spectrum require that 
the envelope is optically thick up to about $\rm{45{\ }\mu}$m. This implies 
that the lack of near infrared flux must be due to a large opacity towards the 
central star, contrary to other post-AGB objects where the lack of near 
infrared flux is due to the lack of dust close to the star 
\citep[e.g.][]{2002A&A...389..547H}. The high mass-loss rate suggested by our 
preliminary model is consistent with this large opacity, and can thus explain 
the lack of near infrared flux. However, more detailed modelling in the form 
of 2D radiative transfer calculations is needed in order to properly constrain
the mass-loss rate of IRAS\,16342. 

The high mass-loss discussed above is only expected to occur during the SW 
phase of the AGB. Combined with the proto-PN nature of IRAS\,16342, we conclude
that IRAS\,16342 must be a very young proto-PN, perhaps the youngest one 
observed so far. The star seems to be right at the transition point between 
the AGB phase (which explains its extreme OH/IR type spectrum) and the 
proto-PN phase. This makes IRAS\,16342 a very interesting object for the study 
of the transition between OH/IR type AGB stars and proto-PN or post-AGB stars.
The young proto-PN nature of IRAS\,16342 is in agreement with SAH99. Still, 
the possibility of a more evolved central star can not be excluded, since the 
$\rm{H_{2}O}$ maser velocities in the lobes (see Sect. 
\ref{sect:introduction}) are more in favour of an A or late type B star. It is
difficult to understand how a later type star can drive such a fast wind. 

\section{The infrared images}

\label{sect:IRimages}

\subsection{Description of the images}

\begin{figure*}
  \caption{The deconvolved L band (ISAAC) and N7.9, N9.8, N2 and Q band
           (TIMMI2) images of IRAS\,16342. The lower right image shows the 
	   division of the N2 band image by the Q band image. Each image 
	   measures 8$''$ $\times$ 8$''$, with north is down and east is 
	   right. The contours indicate levels of equal surface brightness. No
	   attempts were made to give absolute flux values (also see Sect. 
	   \ref{sect:timmi2data} and \ref{sect:IRimages}). In order to enhance
	   the contrast of the images they were multiplied with a power law 
	   (of power 4). More details and a discussion of the images are given
	   in Sect. \ref{sect:IRimages}.}
  \centering
  \hspace{0.5cm}\rotatebox{270}{\resizebox{!}{16.8cm}{\includegraphics{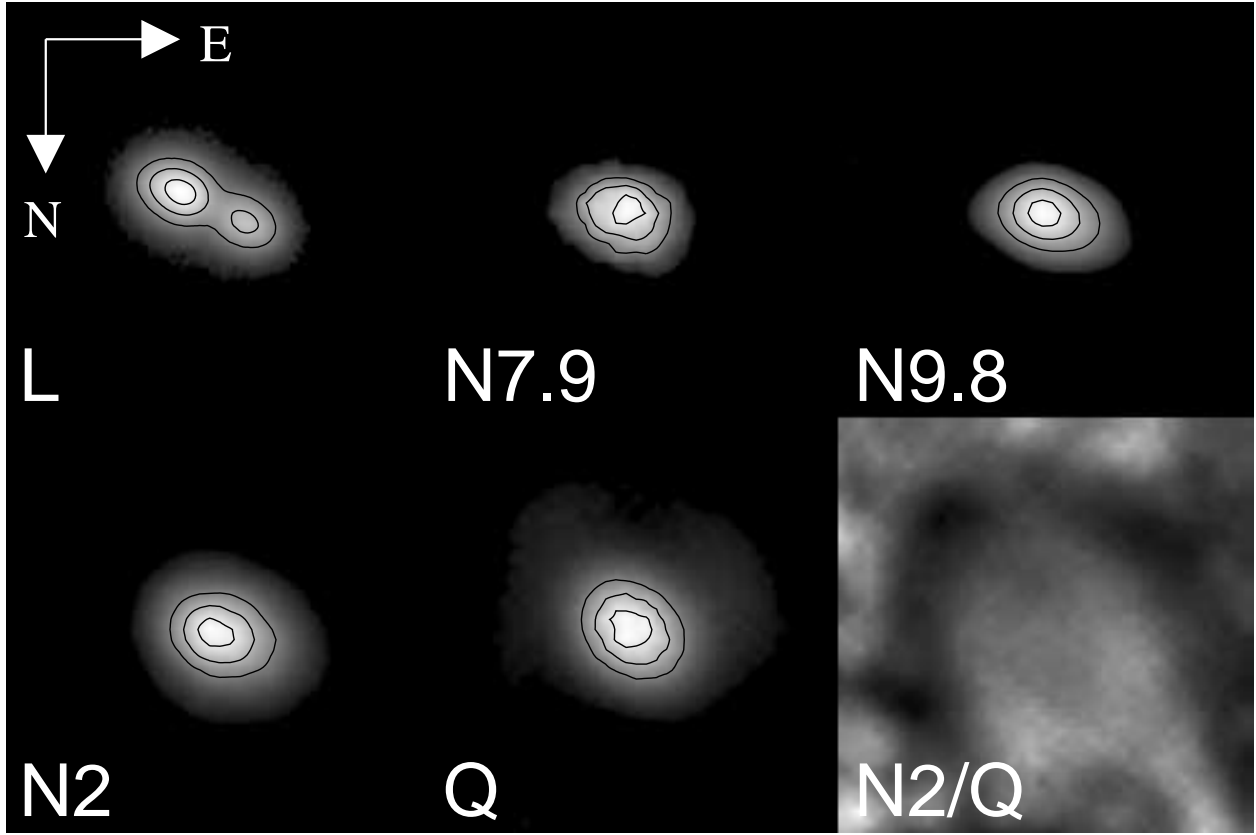}}}
  \label{fig:isaactimmi2images}
\end{figure*}

   \begin{table}[t]
      \caption[]{Size $d$ and position angle $P.A.$ of the bipolar nebula in
		 IRAS\,16342 for different bands/wavelengths. As a function of 
		 wavelength, the size remains approximately the same, while 
		 the position angle increases. For completeness we have also 
		 listed the V band values, taken from SAH99.
		}
         \label{tab:sizesandpas}
     $$ 
         \begin{array}{p{0.1\linewidth}lll}
            \hline
            \noalign{\smallskip}
	    Band&\lambda_{0} ({\mu}m)&\rm{d} ('')&P.A. ({\ }^{\circ})\\
            \noalign{\smallskip}
            \hline
            \noalign{\smallskip}
	    V   &   0.55  &2.6\pm0.3&71\pm2\\
	    L   &   3.78  &3.2\pm0.2&64\pm1\\
	    N7.9&    7.9  &1.3\pm0.3&74~:  \\
	    N9.8&    9.8  &1.8\pm0.3&76\pm2\\
	    N2  &   10.6  &1.4\pm0.3&51\pm2\\
	    Q   &   20.0  &1.4\pm0.3&51\pm2\\
            \noalign{\smallskip}
            \hline
         \end{array}
     $$ 
   \end{table}

In Fig. \ref{fig:isaactimmi2images} we present the deconvolved ISAAC and 
TIMMI2 images of IRAS\,16342. The images were taken in the L, N7.9, N9.8, N2 
and Q bands. The central wavelengths of these bands, $\lambda_{0}$, are given 
in Table \ref{tab:sizesandpas}. The lower right image shows the division of the
N2 band image by the Q band image. 

IRAS\,16342 reveals a bipolar morphology in the L band. The bipolar nature
is similar to that seen in the optical images taken with the HST by 
SAH99. In both cases we see two lobes, surrounded by a faint halo that follows
the contours of the lobes. The TIMMI2 images reveal an ellipsoidal morphology,
which {\bf may} represent the bipolar nebula in the ISAAC and HST images, but 
at lower spatial resolution. {\bf Some caution is needed here however. The 
optical seeing during the observations with ISAAC and TIMMI2 were $1.2-2.0''$ 
and $0.9-2.0''$ respectively (see Sect. \ref{sect:isaacdata} and Sect. 
\ref{sect:timmi2data}). In general, the seeing is better at longer wavelengths 
under the same optical seeing. Therefore we might also expect the TIMMI2 
images to have an equal or even better spatial resolution than the ISAAC image,
in which case differences seen between the images are intrinsic to IRAS 16342.}

For each image, the size $d$ of the nebula along its major axis is listed in 
Table \ref{tab:sizesandpas}, together with $P.A.$, its position angle. We 
define the position angle as the angle between the major axis of the bipolar 
nebula and the north-south line, where ${0}^{\circ}$ represents north and 
${90}^{\circ}$ represents east. Contrary to the size of the nebula, $P.A.$ 
changes significantly as a function of wavelength, showing a decrease with 
increasing wavelength. For the TIMMI2 images, both $d$ and $P.A.$ were 
measured by fitting two dimensional Gaussians to the images. The Full Width at
Half Maximum (FWHM) of the Gaussian along its major axis and its tilt angle 
(i.e. $P.A.-90^{\circ}$), were used to obtain d and $P.A.$ respectively. For 
the ISAAC image we estimated $d$ and $P.A.$ by eye, while for the HST image we
either used values given by SAH99 or derived these values from their data.

The N7.9 band and the Q band reveal (sub)structure additional to the
ellipsoidal structure mentioned in the previous paragraph. The N7.9
band shows some extra extended emission in the north-east direction. Also, 
compared to the other bands, the morphology in the N7.9 band seems to be less 
smooth, with more spatial brightness variations. However, due to the low 
signal-to-noise quality of the data it is difficult to determine if the 
structures are real. Therefore, they should be treated with caution. The Q 
band image reveals extended emission, or a halo, surrounding the central 
ellips. The halo is roughly spherically symmetric, with a gap in the north 
eastern direction, and approximately 6$''$ in diameter, i.e. twice the size of
the bipolar nebula. The halo can also be seen in the lower right part of Fig. 
\ref{fig:isaactimmi2images}, displaying the division of the N2 band and the Q 
band images. The division shows that the halo is not uniformly bright.

In general our infrared images nicely agree with those made by 
\citet{1999ApJS..122..221M} (also see Sect. \ref{sect:introduction}). The 
halo we claim to see at $\rm{20.0{\ }\mu}$m has not been reported by 
\citet{1999ApJS..122..221M} however. It will therefore be interesting to make 
additional observations at $\rm{20.0{\ }\mu}$m in the future.

\subsection{The morphology of IRAS\,16342}

\subsubsection{The lobes}

\label{sect:thelobes}

Given the similarities with the HST image, it seems natural to explain the 
bipolar nebula in the L band as due to reflection as well. Using this 
assumption, we will show in Sect. \ref{sect:grainsizelimits} that the dust 
particles in the lobes must be small (${\leq}0.1{\ }\rm{\mu}$m). If scattering
by small particles is indeed responsible for the presence of the bipolar 
nebula in the ISAAC image at $3.78\,{\mu}$m, a very red bright source is 
needed to illuminate the nebula, since small dust particles usually scatter 
poorly at wavelengths large compared to their radii. This very red source may 
either be the central star of IRAS\,16342 or the inner regions of the torus. 
If it is the central star, this suggests that IRAS\,16342 only recently 
entered the post-AGB phase, in agreement with our results from Sect. 
\ref{sect:masslosshistory}.

An alternative to the scattering scenario in the L band might be that the
observed light is caused by thermal dust emission. This is unlikely however. 
The dust temperature in the lobes at a distance $r$ from the central star, 
$T_{\rm{d}}$, may be represented by 
$T_{\rm{d}}=T_{\rm{eff}}{(r/R_{\star})}^{-0.4}$. This expression assumes that 
the lobes are optically thin for photospheric radiation and that the dust 
emission efficiency scales as $Q(\lambda)\propto{\lambda}^{-1}$. Using Wiens 
law, we need $T_{\rm{d}}{\sim}\,650{\ }\rm{K}$ to emit efficiently at 
$3.78\,{\mu}$m. With $T_{\rm{eff}}{\sim}\,2670{\ }\rm{K}$ and 
$R_{\star}=372{\ }\rm{R_{\odot}}$ (also see Sect. \ref{sect:masslosshistory}) 
we find that the radius of the region that can be heated to this temperature 
is $r{\sim}\,60{\ }\rm{AU}$. Using the distance of 2 kpc given by SAH99 and an
angular diameter of 3$''$, the observed radius of the lobes is 3000 AU. This 
is 50 times larger than the expected typical size of the emitting region 
derived above. We therefore exclude thermal dust emission as the source of 
radiation in the ISAAC image. Non equilibrium emission from small grains is 
also unlikely, since this requires the presence of UV photons 
\citep{1978A&A....66..169A} and these are not likely to be present in the 
system. However, it can not be ruled out completely.

The behaviour of $P.A.$ as a function of wavelength suggests that (as 
expected) the dominant source of radiation changes from scattering to thermal 
emission. At the shortest wavelengths the source is closely aligned in the 
direction of the bipolar scattering nebulae seen in the HST images; at longer
wavelengths the source tends to align more with the torus described in Sect. 
\ref{sect:introduction}, which is expected to be dominated by thermal 
emission. Still, even up to $\rm{20{\ }\mu}$m the elongation is approximately 
along the major axis of the nebula seen with HST and ISAAC, suggesting that
the torus must be very cold ($<150{\ }\rm{K}$). This is in agreement with the 
very red ISO spectrum, which is most likely dominated by thermal emission from
the material in the torus.

\subsubsection{The halo}

We expect the halo in the Q band image to be due to a previous, more 
spherical, episode of mass-loss on the AGB. Given the size of the halo (6$''$)
and the bipolar nebula (3$''$) and assuming an outflow velocity of 
$\rm{15{\ }km{\ }s^{-1}}$ during the AGB phase, and a distance of 2 kpc 
(SAH99), we find that the mass in the halo was ejected between approximately 
950 and 1900 years ago. The Q band image therefore suggests that the 
transition from a (more) spherical to a more axial symmetric form of mass-loss
for IRAS\,16342 only occured near the very end of the AGB. We speculate that 
the non uniform brightness distribution may indicate that there are density 
enhancements in the halo, but this remains questionable. More observations are
required to investigate this possibility.

\subsection{Limits on grain sizes}

\label{sect:grainsizelimits}

Using a simple model, we can put constraints on the particle size of the 
dust in both the torus and the lobes. Now we will explain the model in
detail and discuss its results. We will show that there are two independent 
methods that allow us to derive the optical depth in the torus along the 
line-of-sight towards the eastern lobe. The first method depends on the 
dust particle size in the lobes, while the second method depends on the 
particle size in the torus. Both methods must yield the same optical depth for
a given wavelength. Applying this constraint and calculating the optical depth
with both methods for three different wavelengths, yields the particle size in
both the lobes and the torus. 

\begin{figure}
  \caption{The assumed morphology for IRAS\,16342 used to determine the dust 
	   particle sizes in both the torus and the lobes. For details and a 
	   discussion see Sect. \ref{sect:grainsizelimits}.}
  \vspace{0.25cm}
  \centering
  \hspace{-1cm}\rotatebox{0}{\resizebox{7.0cm}{!}{\includegraphics{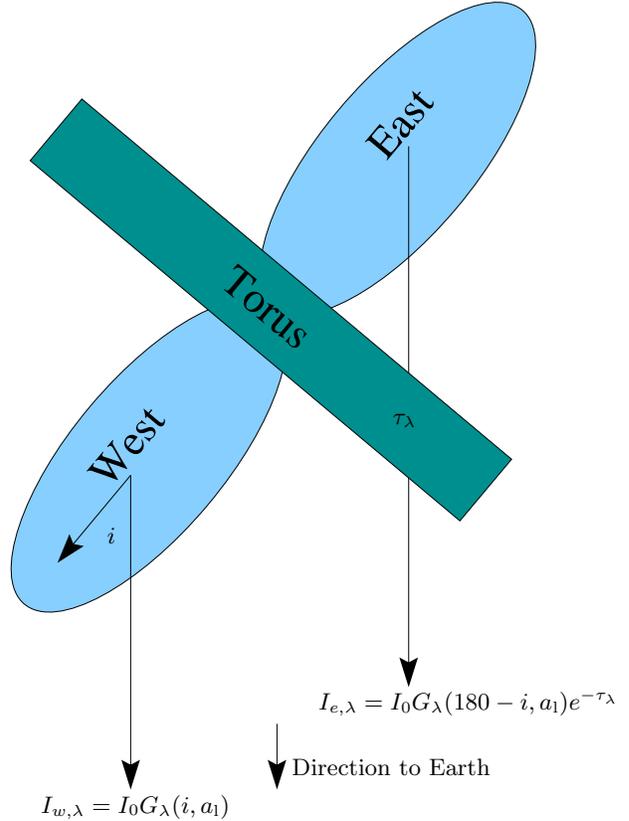}}}\\
\begin{picture}(1,0)
\put(-100,3){$I_{w,\lambda}=I_0 G_\lambda(i,a_{\rm{l}})$}
\put(33,150){${\tau}_{\lambda}$}
\put(5,42){$I_{e,\lambda}=I_0 G_\lambda(180-i,a_{\rm{l}})e^{-\tau_\lambda}$}
\put(-5,17){Direction to Earth}
\put(-75,105){$i$}
\end{picture}
  \label{fig:explainmodel}
\end{figure}

Our model makes use of the following assumptions. The eastern lobe in the
HST and ISAAC images is significantly redder than the western lobe. In order 
to explain this additional reddening, we assume the eastern lobe to be located
completely behind the torus, and the western lobe to be positioned completely 
in front of the torus (see Fig. \ref{fig:explainmodel}). The torus acts as a 
screen that causes additional reddening of the eastern lobe with respect to 
the western lobe. We assume that scattering is the dominant source of emission
in the lobes, and that both scattering lobes are intrinsically equally bright.
Even in the absence of the (extinction due to the) torus, the lobes will not 
appear equally bright as seen from Earth, due to the orientation of the 
system. The orientation causes the eastern lobe to be dominated by back 
scattering, while the western lobe is dominated by forward scattering. 
As a result, the eastern lobe will appear less bright than the western lobe. 
Interstellar reddening is assumed to be the same for both lobes, and 
scattering of stellar radiation within the torus is assumed to be unimportant 
(due to its large radial optical depth).

In the first method to derive the optical depth in the torus we measure the 
intensities of the eastern and western lobes, $I_{e,\lambda}$ and 
$I_{w,\lambda}$, as seen from Earth (see Fig. \ref{fig:explainmodel}). These 
are given by 

\begin{equation}
I_{e,\lambda}=I_0 G_\lambda(180-i,a_{\rm{l}})e^{-\tau_\lambda}
\label{eq:east}
\end{equation}
\begin{equation}
I_{w,\lambda}=I_0 G_\lambda(i,a_{\rm{l}})
\label{eq:west}
\end{equation}
respectively, where $I_0$ is the intrinsic intensity of each lobe. $a_{\rm{l}}$
is the characteristic particle size in the lobes, while 
$G_\lambda(\theta,a_{\rm{l}})$ is the scattering phase function of a particle 
with radius $a_{\rm{l}}$ irradiated at wavelength $\lambda$ and scattering at 
an angle $\theta$. The phase function was calculated for size distributions
of spherical grains using Mie theory \citep{Bohren&Huffman}. Finally, $i$ 
is the inclination angle of the system, defined as the angle between the major
axis of the nebula and the line-of-sight. We take $i=40^{\circ}$ (SAH99). 
$\tau_\lambda$ is the optical depth of the torus along the line-of-sight 
towards the eastern lobe. From Eq. (\ref{eq:east}) and (\ref{eq:west}) we find

\begin{equation}
{\tau}_{\lambda}=\ln\bigg(\frac{{I_{w,\lambda}}{G_{\lambda}(180-i,a_{\rm{l}})}}{{I_{e,\lambda}}{G_{\lambda}(i,a_{\rm{l}})}}\bigg)
\label{eq:tau1}
\end{equation}
This yields $\tau_\lambda$ for different values of $a_{\rm{l}}$.

The intensity ratio $I_{w,\lambda}/I_{e,\lambda}$ was determined from the 
magnitude difference between the lobes, ${\Delta}m$, which was either given by
SAH99 or we measured it ourselves. This implies that we replaced the intensity
ratios by flux ratios. The values for the magnitude differences are listed in 
Table \ref{tab:intensities}, together with the corresponding intensity ratios.

We describe the grain size distribution used in the calculation of 
$G_\lambda(\theta,a_{\rm{l}})$ by a power law $n(a){\propto}a^{-m}$ with 
$m=-3$. This value was chosen for calculation conveniences. Differences with 
models using the more generally adopted ISM value of -3.5 
\citep[e.g.][]{1977ApJ...217..425M} are expected to be small. The effective 
width of the size distribution was kept fixed. The grains in the lobes were 
assumed to be composed of silicates, for which we used the optical constants 
of `cold' silicates derived by \citet{1999MNRAS.304..389S}. Using the optical 
constants of `warm' silicates derived by \citet{1999MNRAS.304..389S} yields 
very similar results (not shown).

In the second method we derive $\tau_\lambda$ from the well known relation

\begin{equation}
{\tau}_{\lambda}=n_{d}{\pi}a_{\rm{t}}^{2}Q^{\rm{ext}}_{\lambda}(a_{\rm{t}})L
\label{eq:tau2}
\end{equation}
where $n_{d}$ is the dust particle number density in a dust column of length
$L$ along the line-of-sight; $a_{\rm{t}}$ and 
$Q^{\rm{ext}}_{\lambda}(a_{\rm{t}})$ are the characteristic radius and
extinction efficiency of the particles in the torus. For the calculation of 
the extinction efficiency we adopt the same dust properties as those assumed 
above for the calculation of the scattering phase functions.

Eq. (\ref{eq:tau1}) and (\ref{eq:tau2}) should yield the same result for 
$\tau_\lambda$ at a given wavelength. Ratios between optical depths are 
independent of $n_{d}$ and $L$ and therefore allow a relatively easy 
derivation of the particle sizes in the lobes and the torus. 

\begin{figure}
  \caption{$\rm{\frac{{\tau}_{0.55}}{{\tau}_{0.80}}}$ against 
	   $\rm{\frac{{\tau}_{0.80}}{{\tau}_{3.78}}}$ for different particle
	   sizes in the torus (labeled with `Q-ratios') and different 
	   particle sizes in the lobes (labeled with `I-ratios') of 
	   IRAS\,16342. Also shown are the positions for the ISM 
	   \citep[based on][]{1979ARA&A..17...73S} and a grey 
	   absorber (indicated by stars). The arrows indicate directions of 
	   increasing particle sizes. For some points (indicated by diamonds) 
	   the maximum particle size of the used size distribution is shown 
	   (in $\rm{\mu}$m). For a detailed explanation and discussion of this 
	   figure see Sect. 
	   \ref{sect:grainsizelimits}.}
  \centering
  \hspace{0cm}\rotatebox{270}{\resizebox{6.5cm}{!}{\includegraphics{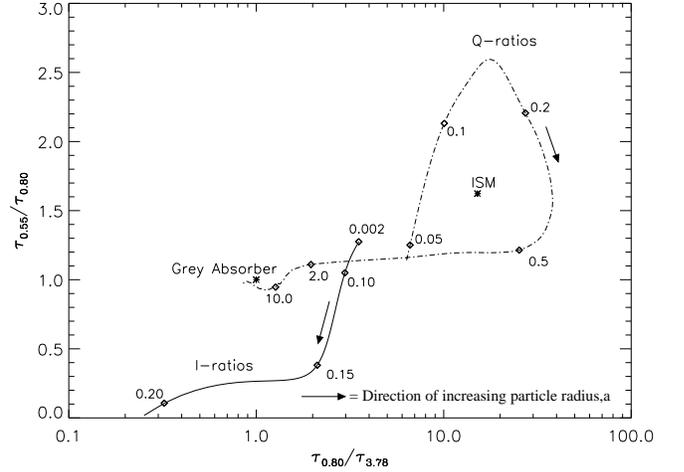}}}
  \label{fig:tautau}
\end{figure}

   \begin{table}[t]
      \caption[]{The intensity ratios, $I_{w,\lambda}/I_{e,\lambda}$, 
		at given wavelengths. This quantity is relevant for the
		calculation of ${\tau}_{\lambda}$ in Eq. (\ref{eq:tau1}).
		Listed are the wavelength; the magnitude difference between
		the lobes; the intensity ratios. The adopted errors on the 
		magnitudes are expected upper limits.
		}
         \label{tab:intensities}
     $$ 
         \begin{array}{ccc}
            \hline
            \noalign{\smallskip}
	    \lambda_{0} ({\mu}m)&{\Delta}m&\frac{I_{w,\lambda}}{I_{e,\lambda}}\\
            \noalign{\smallskip}
            \hline
            \noalign{\smallskip}
	    0.55  &2.8\pm0.1&13.2{\pm}0.5 \\
	    0.79  &2.2\pm0.1& 7.6{\pm}0.3 \\
	    3.78  &0.64\pm0.1& 1.80{\pm}0.07 \\
            \noalign{\smallskip}
            \hline
         \end{array}
     $$ 
   \end{table}

Using Eq. (\ref{eq:tau1}) and the appropriate intensity ratios (see Table 
\ref{tab:intensities}) and phase functions, we measured ${\tau}_{\lambda}$ at 
$0.55$, $0.79$ and $\rm{3.78{\ }\mu}$m for different assumptions for the 
characteristic particle size in the lobes, $a_{\rm{l}}$. In Fig. 
\ref{fig:tautau} $\frac{{\tau}_{0.55}}{{\tau}_{0.80}}$ is plotted versus 
$\frac{{\tau}_{0.80}}{{\tau}_{3.78}}$ with $a_{\rm{l}}$ as a free parameter 
(see line indicated with `I-ratios'). We also calculated 
$Q^{\rm{ext}}_{\lambda}(a_{\rm{t}})$ at the above wavelengths for different 
values of the characteristic particle size in the torus, $a_{\rm{t}}$. Since 
${\tau}_{\lambda}{\propto}Q^{\rm{ext}}_{\lambda}(a_{\rm{t}})$, we must have 

\begin{equation}
\frac{{\tau}_{0.55}}{{\tau}_{0.80}}=\frac{{Q^{\rm{ext}}_{0.55}}(a_{\rm{t}})}{{Q^{\rm{ext}}_{0.80}}(a_{\rm{t}})}
\label{eq:ratios1}
\end{equation}
and

\begin{equation}
\frac{{\tau}_{0.80}}{{\tau}_{3.78}}=\frac{{Q^{\rm{ext}}_{0.80}}(a_{\rm{t}})}{{Q^{\rm{ext}}_{3.78}}(a_{\rm{t}})}
\label{eq:ratios2}
\end{equation}
independent of $n_{d}$ and $L$. In Fig. \ref{fig:tautau} we again plot 
$\frac{{\tau}_{0.55}}{{\tau}_{0.80}}$ versus 
$\frac{{\tau}_{0.80}}{{\tau}_{3.78}}$, this time with $a_{\rm{t}}$ as a free 
parameter (see line indicated with `Q-ratios'). The intersection of the two 
curves provides a consistent set for $a_{\rm{l}}$ and $a_{\rm{t}}$.

For the lobes we find that we have a size distribution with a maximum 
particle size $a^{\rm{max}}_{\rm{l}}=0.09{\pm}0.02{\ }\rm{\mu}$m, while for 
the torus we find $a^{\rm{max}}_{\rm{t}}=1.3{\pm}0.2{\ }\rm{\mu}$m. Both 
values are independent of the adopted effective width of the size 
distribution, which we take to be the same for particles in both the torus and
the lobes. A test with amorphous olivines instead of the cold `Suh' silicates 
yielded a similar result. The errors on the particle sizes are derived from an
error analysis on Eq. (\ref{eq:tau1}) taking the errors on ${\Delta}m$ and 
$I_{w,\lambda}/I_{e,\lambda}$ into account for each wavelength (see Table 
\ref{tab:intensities}). We are interested in the maximum particle size of the 
distribution instead of the typical (i.e. effective) particle size. This is 
because we want to set upper limits on the particle sizes found in the lobes 
and torus. The typical size simply scales linearly with the maximum particle 
size.

We thus derive that the maximum particle size in the torus is about an order 
of magnitude larger than the maximum particle size in the lobes. In an AGB 
outflow, the particle size of the grains formed in the outflow is proportional
to the density of the envelope at the dust condensation radius 
\citep{1992ThesisCarsten}. Larger densities provide larger grains. 
The spatial distribution of large particles in a torus geometry suggests that
this material is expelled from the AGB star in an aspherical high density wind.
In a natural way, this accounts for the presence of large grains in the torus
and small grains in the lobes. However, the bipolar morphology of IRAS\,16342
may also have originated from a later phase, i.e. after the AGB. In this
case, the AGB wind may have been spherical, but due to the interaction of this
wind with the (more recent) fast bipolar molecular outflow, the morphology may
have changed into the observed torus like spatial distribution. The interaction
of these two flows may have caused a shattering of the dust particles in the 
lobes into smaller ones, while leaving the grains in the torus intact. 
Regardless of the correct scenario, the presence of the bipolar nebula in 
IRAS\,16342 at optical wavelengths suggests that the mass-loss rate near the 
central star very recently must have dropped, otherwise the optical photons in
this nebula would not be able to escape. Obviously the star has already 
managed to create a bipolar cavity while the optical depth of the dust shell 
in the equatorial torus is still huge.

The derived particle size for the torus is large compared to the typical ISM 
values, which are of the order of $a\sim{\,}0.01$ to $\sim{\,}0.2\rm{\mu}$m 
\citep[e.g.][]{1977ApJ...217..425M}, but in agreement with other objects for
which large mass-loss rates are expected. Examples of such sources are the 
binary system AFGL 4106 \citep{1999A&A...350..163M} and the post-AGB star 
HD161796 \citep{2002A&A...389..547H}. The large grain size in the torus 
explains why the dust particles in the torus of IRAS\,16342 are grey compared 
to the ISM.

We finally note that when the emission in the ISAAC image at 
$\rm{3.78{\ }\mu}$m is caused by non equilibrium emission from small
dust grains, as was considered in Sect. \ref{sect:thelobes}, the
resulting particle sizes in the lobes and torus are very similar to
those derived above. The possibility of non equilibrium emission by
small grains at $\rm{3.78{\ }\mu}$m can simply be modeled by leaving 
${F_{3.78}(180-i,a_{\rm{l}})}$ and ${F_{3.78}(i,a_{\rm{l}})}$ out 
of equation \ref{eq:tau1} for ${\tau}_{3.78}$, i.e. by assuming that at 
$\rm{3.78{\ }\mu}$m there is no scattering.

\section{Conclusions}

The results of our study can be summarised as follows:

   \begin{enumerate}
        \item The ISO spectrum of IRAS\,16342 is that of an extremely
	reddened OH/IR star. In addition to amorphous silicate 
	absorption features at $10$ and $\rm{20{\ }\mu}$m, the spectrum 
	shows crystalline silicate features in absorption at wavelengths up to
	$\rm{45{\ }\mu}$m. Only at longer wavelengths the crystalline features
	are in emission. No other OH/IR star is known to show this behaviour. 
	We estimate the mass-loss rate of IRAS\,16342 may be as large as 
	$\rm{{10}^{-3}{\ }M_{\odot}{yr}^{-1}}$. This high rate explains the 
	lack of flux in the near-IR as well as the peak wavelength in the SED.
	\item We identify forsterite ($\rm{Mg_{2}SiO_{4}}$), diopside 
	($\rm{MgCaSi_{2}O_{6}}$), and possibly clino-enstatite 
	($\rm{MgSiO_{3}}$) in the spectrum. Besides these crystalline 
	silicates, also crystalline water ice is detected. All these materials
	are also found around other (high $\dot{M}$), oxygen rich, evolved 
	stars. At $\rm{48{\ }\mu}$m a (to date) unidentified feature is seen.
        \item IRAS\,16342 must be a very young proto-PN, perhaps the youngest
        one observed until present. The central star seems to be right at the 
	transition point between the AGB phase and the proto-PN phase. 
        \item The deconvolved L band ISAAC image of IRAS\,16342 at 
	$\rm{3.78{\ }\mu}$m reveals a bipolar scattering nebula. Both lobes of
	the nebula are separated by a dark equatorial waist, or torus. 
	This morphology is consistent with that found in HST images. The 
	presence of the scattering nebula in the L band suggests that the 
	central star of IRAS\,16342 may be very red, which would be in 
	agreement with the young proto-PN nature of the system. The TIMMI2 Q 
	band image reveals thermal emission from an extended, roughly 
	spherical halo of $\sim\,6''$ in diameter. We expect the halo to be 
	due to a previous, more spherical, phase of mass-loss on the AGB. This
	halo material was ejected between approximately 950 and 1900 years ago.
        \item TIMMI2 images at wavelengths intermediate to the L and Q 
	band reveal an ellipsoidal morphology, which {\bf may represent} 
	the bipolar nebula but at lower spatial resolution. The position angle
	of the nebula changes as a function of wavelength, {\bf suggesting} 
	that the nature of the radiation changes from scattered light to 
	thermal emission (by cold material). Still, even up to 
	$\rm{20{\ }\mu}$m the elongation is approximately along the major axis
	of the nebula seen in the optical and near-IR, suggesting that the 
	equatorial torus must be very cold. This is in agreement with the very
	red ISO spectrum, which is most likely dominated by thermal emission 
	from this torus.
	\item Using a simple model for the geometry of IRAS\,16342, we 
	calculate the dust particle sizes in both the torus and the lobes. 
	The grains in the lobes have a maximum particle size 
	$a^{\rm{max}}_{\rm{l}}=0.09{\ }\rm{\mu}$m, while for the torus we find
	$a^{\rm{max}}_{\rm{t}}=1.3{\ }\rm{\mu}$m. The relatively large torus 
	grains suggest that the bipolar morphology of IRAS 16342 may already 
	have been created during the AGB phase. 
   \end{enumerate}

\begin{acknowledgements}
C. Dijkstra wishes to thank Sacha Hony for his help on the data reduction of 
the TIMMI2 data. We are grateful for the support by the staff of the ESO 3.6m 
telescope. 
\end{acknowledgements}

\bibliographystyle{aa}
\bibliography{h4019.bib}

\begin{thebibliography}{31}
\expandafter\ifx\csname natexlab\endcsname\relax\def\natexlab#1{#1}\fi

\bibitem[{{Andriesse}(1978)}]{1978A&A....66..169A}
{Andriesse}, C.~D. 1978, \aap, 66, 169

\bibitem[{Bohren \& Huffman(1983)}]{Bohren&Huffman}
Bohren, C.~F. \& Huffman, D.~R. 1983, Absorption and Scattering of Light by
  Small Particles (New York: Wiley)

\bibitem[{{Bouwman}(2001)}]{2001PhDT..........B}
{Bouwman}, J. 2001, Ph.D.~Thesis University of Amsterdam

\bibitem[{{Clegg} {et~al.}(1996){Clegg}, {Ade}, {Armand}, {Baluteau}, {Barlow},
  {Buckley}, {Berges}, {Burgdorf}, {Caux}, {Ceccarelli}, {Cerulli}, {Church},
  {Cotin}, {Cox}, {Cruvellier}, {Culhane}, {Davis}, {di Giorgio}, {Diplock},
  {Drummond}, {Emery}, {Ewart}, {Fischer}, {Furniss}, {Glencross},
  {Greenhouse}, {Griffin}, {Gry}, {Harwood}, {Hazell}, {Joubert}, {King},
  {Lim}, {Liseau}, {Long}, {Lorenzetti}, {Molinari}, {Murray}, {Naylor},
  {Nisini}, {Norman}, {Omont}, {Orfei}, {Patrick}, {Pequignot}, {Pouliquen},
  {Price}, {Nguyen-Q-Rieu}, {Rogers}, {Robinson}, {Saisse}, {Saraceno},
  {Serra}, {Sidher}, {Smith}, {Smith}, {Spinoglio}, {Swinyard}, {Texier},
  {Towlson}, {Trams}, {Unger}, \& {White}}]{1996A&A...315L..38C}
{Clegg}, P.~E., {Ade}, P.~A.~R., {Armand}, C., {et~al.} 1996, \aap, 315, L38

\bibitem[{{de Graauw} {et~al.}(1996){de Graauw}, {Haser}, {Beintema},
  {Roelfsema}, {van Agthoven}, {Barl}, {Bauer}, {Bekenkamp}, {Boonstra},
  {Boxhoorn}, {Cote}, {de Groene}, {van Dijkhuizen}, {Drapatz}, {Evers},
  {Feuchtgruber}, {Frericks}, {Genzel}, {Haerendel}, {Heras}, {van der Hucht},
  {van der Hulst}, {Huygen}, {Jacobs}, {Jakob}, {Kamperman}, {Katterloher},
  {Kester}, {Kunze}, {Kussendrager}, {Lahuis}, {Lamers}, {Leech}, {van der
  Lei}, {van der Linden}, {Luinge}, {Lutz}, {Melzner}, {Morris}, {van Nguyen},
  {Ploeger}, {Price}, {Salama}, {Schaeidt}, {Sijm}, {Smoorenburg}, {Spakman},
  {Spoon}, {Steinmayer}, {Stoecker}, {Valentijn}, {Vandenbussche}, {Visser},
  {Waelkens}, {Waters}, {Wensink}, {Wesselius}, {Wiezorrek}, {Wieprecht},
  {Wijnbergen}, {Wildeman}, \& {Young}}]{1996A&A...315L..49D}
{de Graauw}, T., {Haser}, L.~N., {Beintema}, D.~A., {et~al.} 1996, \aap, 315,
  L49

\bibitem[{{Devillard}(1997)}]{1997TheMessenger...87S}
{Devillard}, N. 1997, The Messenger, 87

\bibitem[{{Dominik}(1992)}]{1992ThesisCarsten}
{Dominik}, C. 1992, Thesis Technischen Universit\"{a}t Berlin

\bibitem[{{Ferrarotti} {et~al.}(2000){Ferrarotti}, {Gail}, {Degiorgi}, \&
  {Ott}}]{2000A&A...357L..13F}
{Ferrarotti}, A., {Gail}, H.-P., {Degiorgi}, L., \& {Ott}, H.~R. 2000, \aap,
  357, L13

\bibitem[{{Guertler} {et~al.}(1996){Guertler}, {Koempe}, \&
  {Henning}}]{1996A&A...305..878G}
{Guertler}, J., {Koempe}, C., \& {Henning}, T. 1996, \aap, 305, 878

\bibitem[{{Hoogzaad} {et~al.}(2002){Hoogzaad}, {Molster}, {Dominik}, {Waters},
  {Barlow}, \& {de Koter}}]{2002A&A...389..547H}
{Hoogzaad}, S.~N., {Molster}, F.~J., {Dominik}, C., {et~al.} 2002, \aap, 389,
  547

\bibitem[{{Kemper} {et~al.}(2002){Kemper}, {de Koter}, {Waters}, {Bouwman}, \&
  {Tielens}}]{2002A&A...384..585K}
{Kemper}, F., {de Koter}, A., {Waters}, L.~B.~F.~M., {Bouwman}, J., \&
  {Tielens}, A.~G.~G.~M. 2002, \aap, 384, 585

\bibitem[{{Koike} {et~al.}(2000){Koike}, {Tsuchiyama}, {Shibai}, {Suto},
  {Tanab{\' e}}, {Chihara}, {Sogawa}, {Mouri}, \&
  {Okada}}]{2000A&A...363.1115K}
{Koike}, C., {Tsuchiyama}, A., {Shibai}, H., {et~al.} 2000, \aap, 363, 1115

\bibitem[{{Koike} {et~al.}(1999){Koike}, {Tsuchiyama}, \&
  {Suto}}]{1999SpectraKoike}
{Koike}, C., {Tsuchiyama}, A., \& {Suto}, H. 1999, Proc. of the 32nd ISAS Lunar
  and Planetary Symposium, 32, 175

\bibitem[{{Likkel} \& {Morris}(1988)}]{1988ApJ...329..914L}
{Likkel}, L. \& {Morris}, M. 1988, \apj, 329, 914

\bibitem[{{Likkel} {et~al.}(1992){Likkel}, {Morris}, \&
  {Maddalena}}]{1992A&A...256..581L}
{Likkel}, L., {Morris}, M., \& {Maddalena}, R.~J. 1992, \aap, 256, 581

\bibitem[{{Mathis} {et~al.}(1977){Mathis}, {Rumpl}, \&
  {Nordsieck}}]{1977ApJ...217..425M}
{Mathis}, J.~S., {Rumpl}, W., \& {Nordsieck}, K.~H. 1977, \apj, 217, 425

\bibitem[{{McGregor}(1994)}]{1994PASP..106..508M}
{McGregor}, P.~J. 1994, \pasp, 106, 508

\bibitem[{{Meixner} {et~al.}(2002){Meixner}, {Ueta}, {Bobrowsky}, \&
  {Speck}}]{2002ApJ...571..936M}
{Meixner}, M., {Ueta}, T., {Bobrowsky}, M., \& {Speck}, A. 2002, \apj, 571, 936

\bibitem[{{Meixner} {et~al.}(1999){Meixner}, {Ueta}, {Dayal}, {Hora}, {Fazio},
  {Hrivnak}, {Skinner}, {Hoffmann}, \& {Deutsch}}]{1999ApJS..122..221M}
{Meixner}, M., {Ueta}, T., {Dayal}, A., {et~al.} 1999, \apjs, 122, 221

\bibitem[{{Molster} {et~al.}(2002{\natexlab{a}}){Molster}, {Waters}, \&
  {Tielens}}]{2002A&A...382..222M}
{Molster}, F.~J., {Waters}, L.~B.~F.~M., \& {Tielens}, A.~G.~G.~M.
  2002{\natexlab{a}}, \aap, 382, 222

\bibitem[{{Molster} {et~al.}(2002{\natexlab{b}}){Molster}, {Waters}, {Tielens},
  \& {Barlow}}]{2002A&A...382..184M}
{Molster}, F.~J., {Waters}, L.~B.~F.~M., {Tielens}, A.~G.~G.~M., \& {Barlow},
  M.~J. 2002{\natexlab{b}}, \aap, 382, 184

\bibitem[{{Molster} {et~al.}(1999){Molster}, {Waters}, {Trams}, {Van Winckel},
  {Decin}, {van Loon}, {J{\" a}ger}, {Henning}, {K{\" a}ufl}, {de Koter}, \&
  {Bouwman}}]{1999A&A...350..163M}
{Molster}, F.~J., {Waters}, L.~B.~F.~M., {Trams}, N.~R., {et~al.} 1999, \aap,
  350, 163

\bibitem[{{Sahai} {et~al.}(1999){Sahai}, {Te Lintel Hekkert}, {Morris},
  {Zijlstra}, \& {Likkel}}]{1999ApJ...514L.115S}
{Sahai}, R., {Te Lintel Hekkert}, P., {Morris}, M., {Zijlstra}, A., \&
  {Likkel}, L. 1999, \apjl, 514, L115

\bibitem[{{Savage} \& {Mathis}(1979)}]{1979ARA&A..17...73S}
{Savage}, B.~D. \& {Mathis}, J.~S. 1979, \araa, 17, 73

\bibitem[{{Smith} {et~al.}(1994){Smith}, {Robinson}, {Hyland}, \&
  {Carpenter}}]{1994MNRAS.271..481S}
{Smith}, R.~G., {Robinson}, G., {Hyland}, A.~R., \& {Carpenter}, G.~L. 1994,
  \mnras, 271, 481

\bibitem[{{Suh}(1999)}]{1999MNRAS.304..389S}
{Suh}, K. 1999, \mnras, 304, 389

\bibitem[{{Sylvester} {et~al.}(1999){Sylvester}, {Kemper}, {Barlow}, {de Jong},
  {Waters}, {Tielens}, \& {Omont}}]{1999A&A...352..587S}
{Sylvester}, R.~J., {Kemper}, F., {Barlow}, M.~J., {et~al.} 1999, \aap, 352,
  587

\bibitem[{{Ueta} {et~al.}(2000){Ueta}, {Meixner}, \&
  {Bobrowsky}}]{2000ApJ...528..861U}
{Ueta}, T., {Meixner}, M., \& {Bobrowsky}, M. 2000, \apj, 528, 861

\bibitem[{{van der Bliek} {et~al.}(1996){van der Bliek}, {Manfroid}, \&
  {Bouchet}}]{1996A&AS..119..547V}
{van der Bliek}, N.~S., {Manfroid}, J., \& {Bouchet}, P. 1996, \aaps, 119, 547

\bibitem[{{Zijlstra} {et~al.}(2001){Zijlstra}, {Chapman}, {te Lintel Hekkert},
  {Likkel}, {Comeron}, {Norris}, {Molster}, \& {Cohen}}]{2001MNRAS.322..280Z}
{Zijlstra}, A.~A., {Chapman}, J.~M., {te Lintel Hekkert}, P., {et~al.} 2001,
  \mnras, 322, 280

\bibitem[{{Zuckerman} \& {Lo}(1987)}]{1987A&A...173..263Z}
{Zuckerman}, B. \& {Lo}, K.~Y. 1987, \aap, 173, 263

\end{thebibliography}

\end{document}